\documentclass[traditabstract]{aa}
\usepackage{natbib}
\usepackage{txfonts}
\usepackage{graphicx}
\usepackage{footnote}
\titlerunning{Limb darkening in red giant stars}
\authorrunning{H.~R. Neilson}

\begin{document}

\title{Spherically-symmetric model stellar atmospheres and limb darkening I: 
limb-darkening laws, gravity-darkening coefficients and angular 
diameter corrections for red giant stars\thanks{Tables 2 --17 are only available in electronic form at the CDS via anonymous ftp to cdsarc-u-strasbg.fr or via http://cdsweb.u-strasbg.fr/cgi-bin/qcat?J/A+A/}}

\author{Hilding R. Neilson\inst{1} \and John B. Lester \inst{2,3}}
\institute{
    Department of Physics \& Astronomy, East Tennessee State University, Box 70652, Johnson City, TN 37614 USA
    \email{neilsonh@etsu.edu} 
 \and 
     Department of Chemical and Physical Sciences, 
     University of Toronto Mississauga 
 \and
     Department of Astronomy \& Astrophysics, University of Toronto \\
     \email{lester@astro.utoronto.ca}
  }

\date{}

\abstract{
Model stellar atmospheres are fundamental tools for understanding 
stellar observations from interferometry, microlensing, eclipsing 
binaries and planetary transits.  However, the calculations also 
include assumptions, such as the geometry of the model.  We use 
intensity profiles computed for both plane-parallel and spherically 
symmetric model atmospheres to determine fitting coefficients in the 
$BVRIHK$, {\it CoRot} and {\it Kepler} wavebands for limb darkening 
using several different fitting laws, for gravity-darkening and for 
interferometric angular diameter corrections.  Comparing predicted 
variables for each geometry, we find that the spherically symmetric 
model geometry leads to different predictions for surface gravities 
$\log g < 3$.  In particular, the most commonly used limb-darkening 
laws produce poor fits to the intensity profiles of spherically 
symmetric model atmospheres, which indicates the need for more 
sophisticated laws.  Angular diameter corrections for spherically 
symmetric models range from 0.67 to 1, compared to the much smaller 
range from 0.95 to 1 for plane-parallel models.
}
\maketitle

\section{Introduction}

Stellar limb darkening is an important tool for interpreting  
     interferometric, microlensing and eclipsing binary observations of 
red giant and supergiant stars. It    also provides critical information
about the temperature structure of a stellar atmosphere 
\citep{Schwarzschild1906} as well as a 
measure of the radial extension of 
an atmosphere \citep{Neilson2012}.

Interferometric observations measure the angular diameter of a star 
as well as the intensity variation across the stellar surface.  Some of the first interferometric 
observations measured only uniform-disk angular diameters, that is the 
angular diameter for a star assumed to have a constant surface brightness
         \citep{Hanbury1974}.  \citet{Wittkowski2004} presented $K$-band 
interferometric observations of the M3 giant $\psi$ Phoenicis with 
measurements of the first and second lobes of the visibility curve, 
which constrain limb darkening.  Unfortunately, these observations were 
not precise enough to distinguish between different model stellar atmospheres.  
Advances in interferometric observations have allowed for observations 
of convective cells in Betelgeuse \citep{Haubois2009} and measurements 
of gravity darkening in Altair \citep{vanBelle2001}.  In terms of model 
stellar atmospheres, \citet{Aufdenberg2005} constrained 
three-dimensional models using observations of Procyon.

Microlensing observations, like interferometry, also probe stellar limb 
darkening, but unlike interferometry, which targets specific nearby stars, microlensing observations are random. \citet{An2002} and \citet{Fields2003}
constrained non-linear limb-darkening relations from microlensing 
observations of a K3 giant and compared them to model stellar 
atmospheres.   They found significant disagreement between the observed and predicted limb darkening relation.  More recently, however, microlensing observations have 
only constrained linear limb-darkening relations for red giant stars 
\citep{Fouque2010, Zub2011}.

Eclipsing binaries and planetary transits provide yet another avenue 
for measuring stellar limb darkening.  In terms of red giant stars, 
there are a number of known eclipsing binary systems, specifically the 
$\zeta$ Aurigae systems that have a K4-5 red giant primary and a 
main-sequence B-type companion. \citet{Eaton2008} fit the orbits for 
several of these systems assuming a simple linear limb-darkening law.  
There is also the potential of observing planets transiting            
red giant stars, which would provide powerful constraints of theories 
of planetary evolution.  Currently, extrasolar planets have been observed 
orbiting dwarf and subgiant stars \citep{Howell2012}, but not giant stars; 
future missions such as PLATO may remedy this \citep{Catala2010}.

These three types of observations are ideal tools for probing stellar 
atmospheres and constraining the         physics employed in numerical 
models.  Likewise, predictions from model stellar atmospheres help 
constrain these types of observations.  Recently, \citet{Sing2010}, 
\citet{Howarth2011a} and \citet{Claret2011} presented limb-darkening laws 
fit to plane-parallel model stellar atmosphere intensity profiles.   Even more recently, \cite{Claret2012, Claret2013} fit limb-darkening laws to spherically-symmetric PHOENIX model stellar atmospheres of cool brown dwarf stars. 
In this work, we study how the assumed geometry of the model stellar atmosphere, 
          plane parallel versus spherically symmetric, affects 
predictions of stellar limb darkening, gravity darkening and 
interferometric angular diameter corrections.  We examine model 
atmospheres spanning the effective temperature and gravity range 
consistent with yellow and red giant and supergiant stars.  Tables of 
limb-darkening and gravity-darkening coefficients, as well as new 
angular diameter corrections are presented as more physically based 
tools for understanding these bright stars.

In Sect.~2 we describe the stellar atmosphere code used in this work, 
as well as the model atmosphere grids computed for both plane-parallel and 
spherically symmetric geometries.       In Sect.~3 limb-darkening 
coefficients are presented for several commonly used limb-darkening 
relations.  We compute gravity-darkening coefficients in Sect.~4 and 
angular diameter corrections in Sect.~5.  Computations in these three 
sections provide insight into how intensity profiles depend on the assumed 
model geometry that can be directly compared to observations. 

\section{Model Stellar Atmospheres}

Model stellar atmospheres form a key foundation of our understanding of 
stars, arguably a great success of 
computational astrophysics.  However, the early success of model 
atmosphere codes transformed them into standard tools, and only in the past decade have 
these codes moved beyond simple plane-parallel, 
local-thermodynamic-equilibrium (LTE) models to full 
three-dimensional, statistical-equilibrium codes that can model non-LTE 
physics as well as stellar convection. Unfortunately, computing power 
is still limited for calculating large-scale model atmosphere grids 
varying stellar gravity, effective temperature, stellar mass and 
composition.

A step toward more realistic geometry is achieved by shifting from one-dimensional plane-parallel model stellar atmosphere 
codes to one-dimensional spherically symmetric codes, which 
        can be used to compute large grids of models atmospheres that 
include physics that is more appropriate to stars where the depth of 
the stellar photosphere is a significant fraction of the stellar radius,
such as evolved giant and supergiant stars and pre-main sequence stars. 
One such code for modeling atmospheres assuming spherically symmetric 
geometry is the \textsc{SAtlas} code \citep{Lester2008}.  This code is based on 
the \textsc{Atlas} code developed by \citet{Kurucz1979}, and continues 
its assumption of local 
thermodynamic and hydrostatic equilibrium.  However, the radiative transfer is computed assuming spherical geometry
         using the \citet{Rybicki1971} version of the 
\citet{Feautrier1964} ray-tracing method, while radiative and convecting equilibrium is enforced using an 
updated version of the \citet{Avrett1963} temperature correction method.  Models computed using this code have been compared to spherically-symmetric \textsc{Phoenix} and MARCS models \citep{Hauschildt1999, Gustafsson2008} and shown to produce similar results \citep{Lester2008, Neilson2008}.

In this work we use the grid of spherical model atmospheres from \citet{Neilson2011}, 
extended in mass up to $M = 20~M_\odot$.  The grid 
assumes solar composition and spans the gravities from $\log g = -1$ to 
$\log g = 3$ in steps of $0.25$, effective temperatures from 
$T_\mathrm{eff} = 3000$ to $8000~$K and masses from $M = 2.5$ to 
$20~M_\odot$ in steps of $2.5~M_\odot$ and includes models with masses 
$M = 0.5$ and $1~M_\odot$.  Surface intensities are computed for each model at 
1000 equally spaced values of $\mu = \cos \theta$, where $\theta$ 
is the angle between the vertical direction and the direction toward 
a distant observer.  Limb-darkening profiles are 
computed for Johnson-Cousins $BVRIHK$-wavebands \citep{Johnson1953, Bessell2005} along with the 
{\it CoRot} \citep{Auvergne2009} and {\it Kepler} \citep{Koch2004} wavebands. Angular 
diameter corrections for interferometric observations,  
gravity-darkening coefficients and various limb-darkening relations are 
computed using these wavelength-integrated intensity profiles. 

\section{Limb-Darkening Laws}

An understanding of stellar limb darkening is required to model the 
properties of interferometric, eclipsing binary-star, microlensing, and 
planetary-transit observations.  As 
these observations become more precise and more accurate, models of stellar limb darkening must also improve.  Limb darkening is typically 
treated as a simple parametrization as a function of $\theta$       
                           \citep[e.g.][]{Fouque2010, Croll2011}, which 
makes fitting the stellar intensity profile much simpler and reduces the 
number of free parameters.                               The most 
common parametrizations are linear and quadratic relations 
\citep{Al-Naimiy1979, vanHamme1993, Diaz1995}, but other suggested 
relations include a four-parameter relation \citep{Claret2000}, 
a square-root relation \citep{Wade1985} as well as exponential and 
logarithmic relations \citep{Claret2000, Claret2003}. 
\begin{figure*}[t]
\begin{center}
\includegraphics[width=0.5\textwidth]{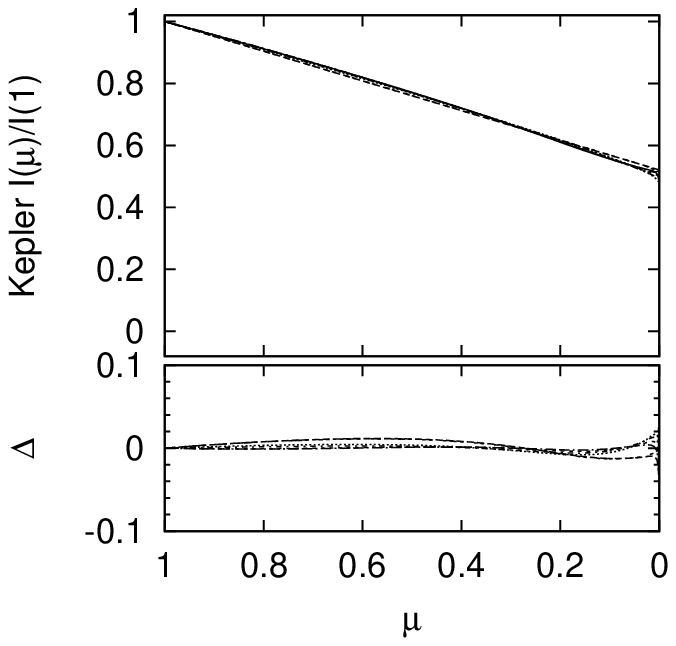}\includegraphics[width=0.5\textwidth]{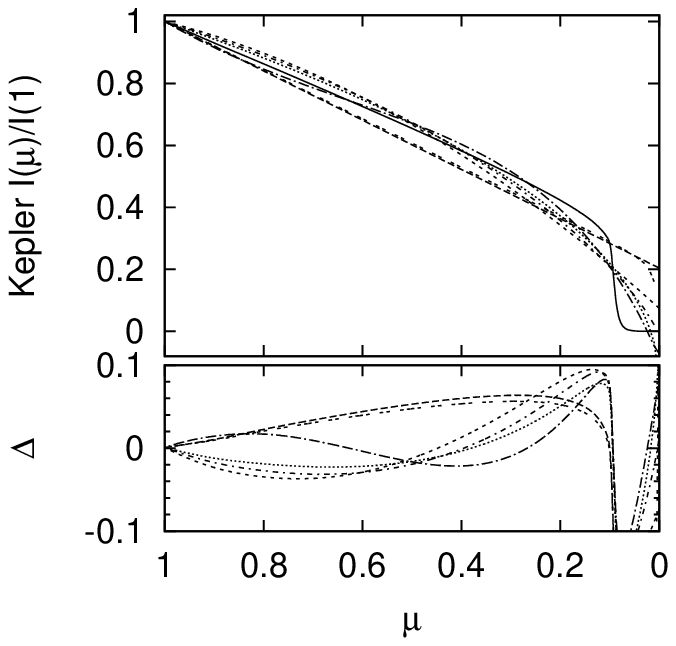}
\end{center}
\caption{{\it Kepler}-band model intensity profiles (black-solid) predicted for both 
plane-parallel (left) and spherically symmetric (right) model stellar 
atmospheres with $T_\mathrm{eff} = 5000~$K, $\log g = 2$ and 
$M = 10~M_\odot$.  Along with the intensity profiles, best-fit linear 
(green-dashed), quadratic (orange-short-dashed), square-root 
(blue-dotted), four-parameter (violet-long-dash-dotted), logarithmic 
(brown-short-dash-dotted), and exponential (grey-double-dash) 
limb-darkening relations are plotted.  Bottom panels show the 
difference, $\Delta \equiv I_\mathrm{model} - I_\mathrm{law}$, between model intensities and best-fit 
limb-darkening laws.}
\label{fig:ld_ex}
\end{figure*}
\begin{figure}[t]
\begin{center}
\includegraphics[width=0.5\textwidth]{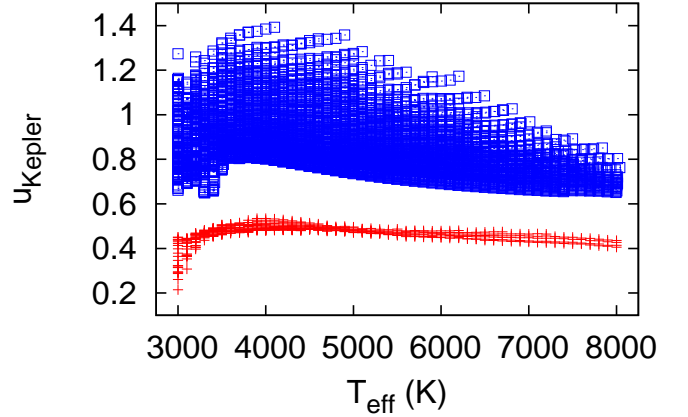}
\end{center}
\caption{The limb-darkening coefficient $u$, used in Eq.~\ref{eq:linear},
 applied to the \textit{Kepler} photometric band. Red crosses are the plane-parallel 
 model stellar atmospheres, and the blue squares are the spherical models. }
\label{fig:ab_l}
\end{figure}

\subsection{Best-Fit Limb-Darkening Laws}

We fit the following limb-darkening relations to the grids of 
plane-parallel and spherically symmetric model stellar atmospheres:
\begin{equation}\label{eq:linear}
\frac{I(\mu)}{I(\mu=1)} = 1 - u(1- \mu) \hfill \makebox{Linear,\hspace{.3cm}}
\end{equation}
\begin{equation}\label{eq:quad}
\frac{I(\mu)}{I(\mu=1)} = 1 - a(1-\mu) - b(1-\mu)^2 \hfill \mbox{Quadratic,\hspace{1 em}}
\end{equation}
\begin{equation}\label{eq:root}
\frac{I(\mu)}{I(\mu=1)} = 1 - c(1-\mu) - d(1-\sqrt{\mu}) \hfill \mbox{Square-Root,\hspace{1 em}}
\end{equation}
\begin{equation}\label{eq:4-p}
\frac{I(\mu)}{I(\mu=1)} = 1 - \sum_{j=1}^{4}f_j(1-\mu^{j/2}) \hfill \mbox{4-Parameter,\hspace{1 em}}
\end{equation}
\begin{equation}\label{eq:exp}
\frac{I(\mu)}{I(\mu=1)} = 1 - g(1-\mu) - h\frac{1}{1-e^{\mu}} \hfill \mbox{Exponential,\hspace{1 em}}
\end{equation}
\begin{equation}\label{eq:ln}
\frac{I(\mu)}{I(\mu=1)} = 1 - m(1-\mu) - n\mu \ln \mu \hfill \mbox{Logarithmic.\hspace{1 em}}
\end{equation}

 We derive the best-fit coefficients for each of the limb-darkening laws using a general least-squares algorithm. This was done using the computed surface intensities  for the $BVRIHK$- and {\it CoRot}- and 
{\it Kepler}-wavebands.   Fig.~\ref{fig:ld_ex} shows the 
{\it Kepler}-band intensity profile and corresponding best-fit 
limb-darkening laws for both spherical and plane-parallel model 
atmospheres with the properties $T_\mathrm{eff} = 5000~$K, $\log g = 2$ 
and $M = 10~M_\odot$ (mass is defined for the spherical model only).   The 
chosen limb-darkening laws all           fit the plane-parallel model 
intensity profiles well.  This is not surprising because plane-parallel model 
atmosphere intensity profiles do not deviate significantly from being 
linear, and a linear term is included in all of the chosen 
limb-darkening laws.  However, spherically symmetric model stellar atmospheres 
                    have      intensity profiles that are significantly 
non-linear, and the best-fit limb-darkening relations for these 
intensity profiles                                                     
match less  well than for the plane-parallel models because of this 
non-linearity.  For the model shown in Fig.~\ref{fig:ld_ex}, 
limb-darkening laws predict intensities that vary by 
$\Delta \equiv I_{\rm{model}} - I_{\rm{law}}  = 0.15$ for the spherical model 
while $\Delta < 0.04$ for the plane-parallel model.  Although 
limb-darkening laws           fit the intensities of plane-parallel model atmospheres 
better than spherically symmetric models, the spherical models are more 
physically realistic, making them the more appropriate choice to use in 
modeling observations.  We explore the uncertainty of the limb-darkening fits later.

\begin{figure*}[t]
\begin{center}
\includegraphics[width=0.5\textwidth]{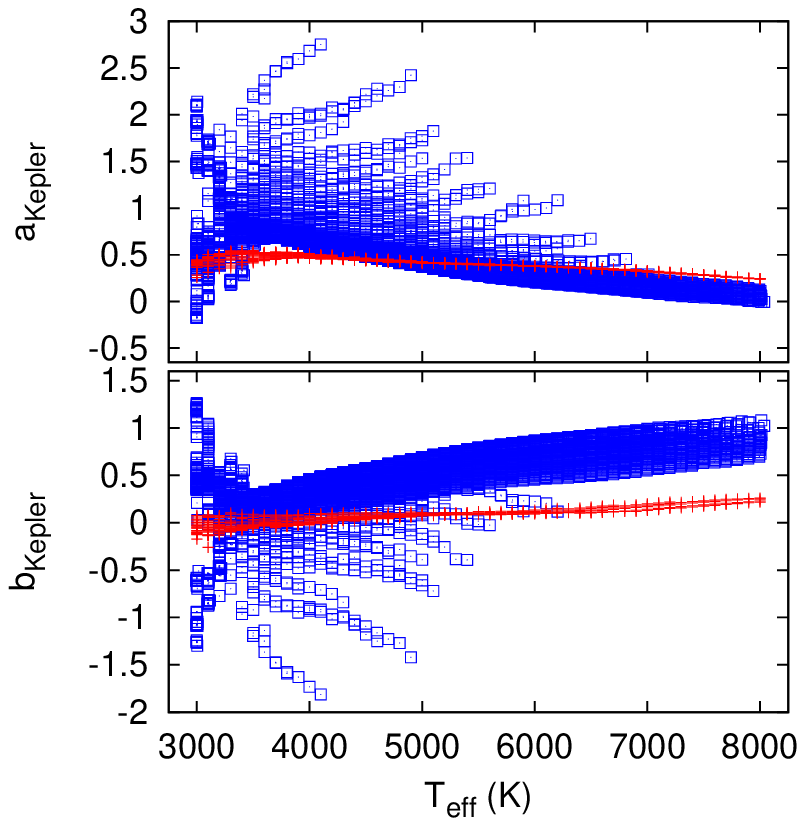}\includegraphics[width=0.5\textwidth]{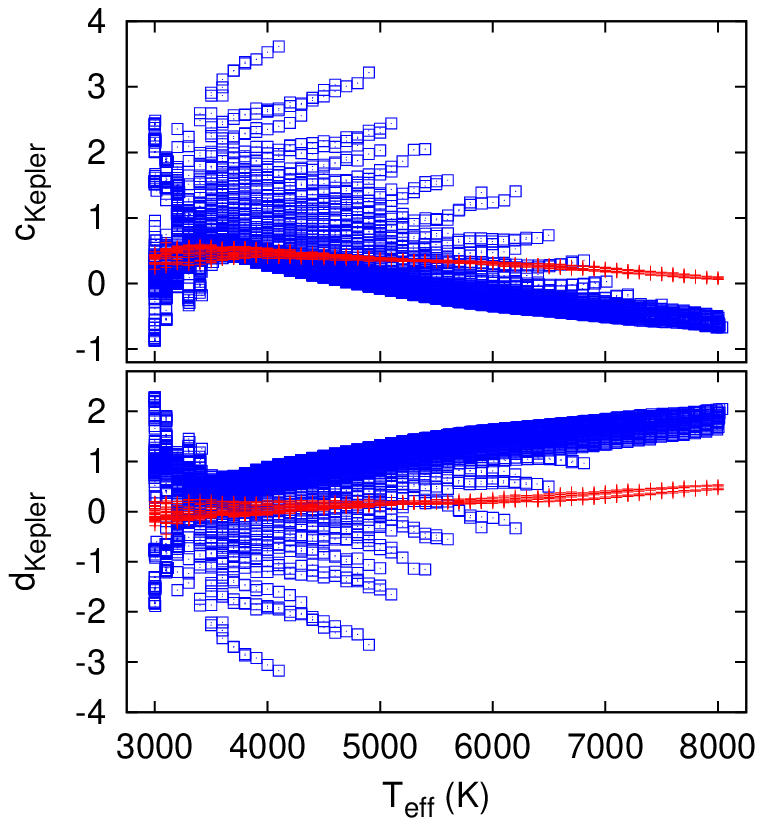}

\end{center}
\caption{Limb-darkening coefficients $a$ and $b$ used in 
 Eq.~\ref{eq:quad} (left panel), and the coefficients $c$ and $d$ used in 
 Eq.~\ref{eq:root} (right panel), all applied to the \textit{Kepler} photometric band.  The symbols 
 have the same meanings as in Fig.~\ref{fig:ab_l}. }
\label{fig:ab_q}
\end{figure*}

\begin{figure*}[t]
\begin{center}
\includegraphics[width=0.5\textwidth]{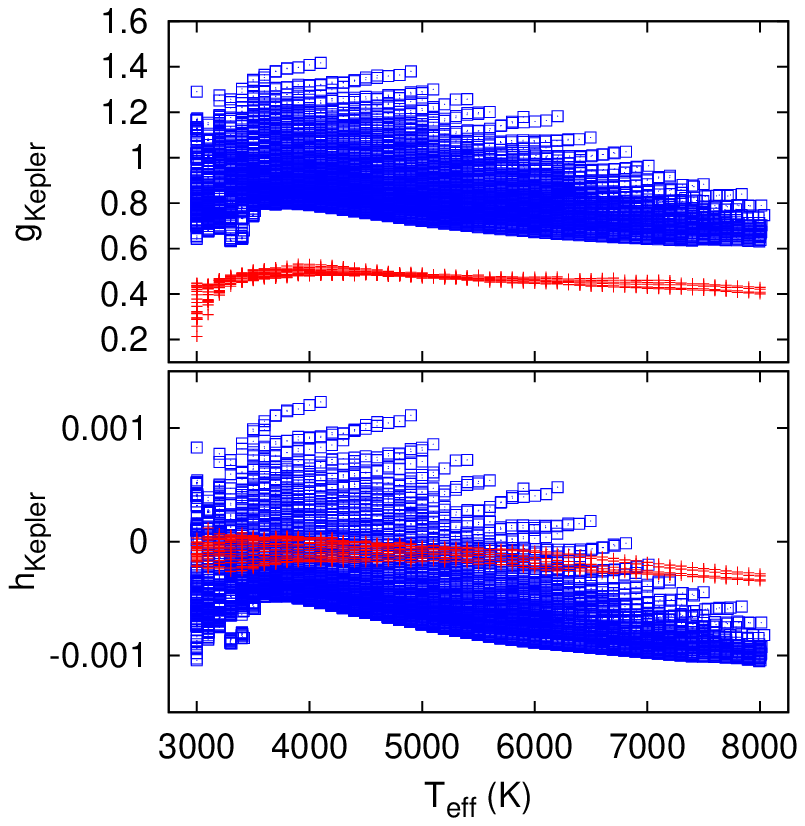}\includegraphics[width=0.5\textwidth]{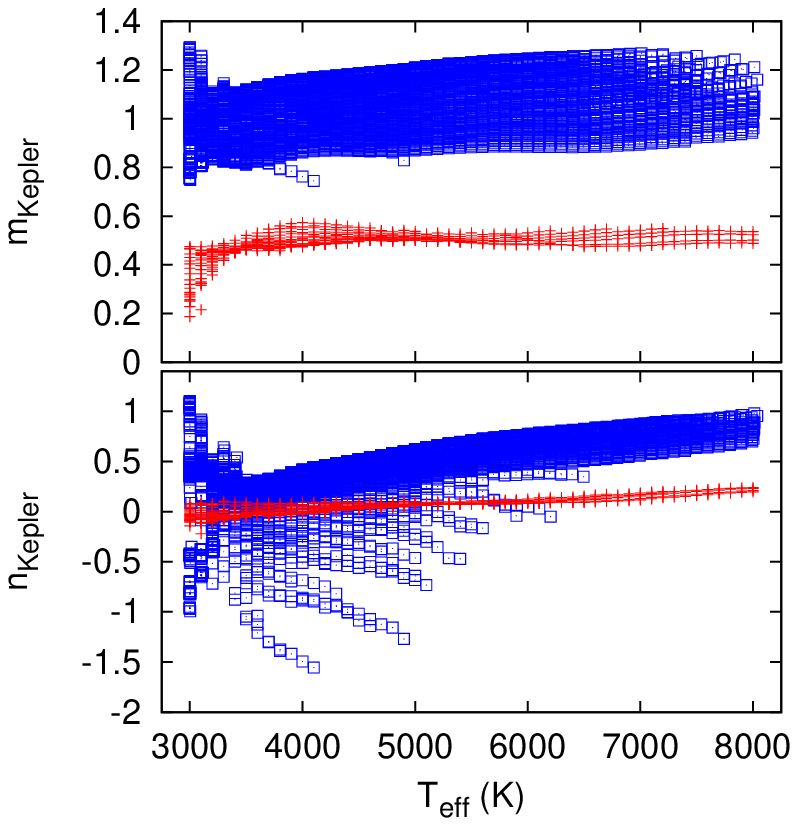}
\end{center}
\caption{Limb-darkening coefficients $g$ and $h$ used in 
 Eq.~\ref{eq:exp} (left panel), and the coefficients $m$ and $n$ used in 
 Eq.~\ref{eq:ln} (right panel), all applied to the \textit{Kepler} photometric band.  The symbols 
 have the same meanings as in Fig.~\ref{fig:ab_l}. }
\label{fig:ab_log}
\end{figure*}

\begin{figure*}[t]
\begin{center}
\includegraphics[width=0.5\textwidth]{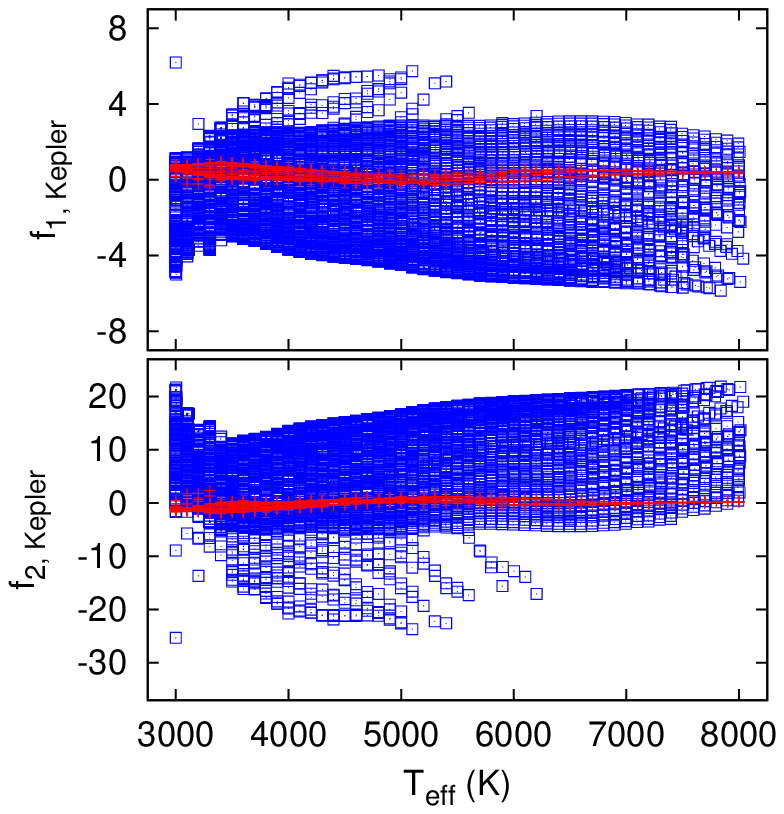}\includegraphics[width=0.5\textwidth]{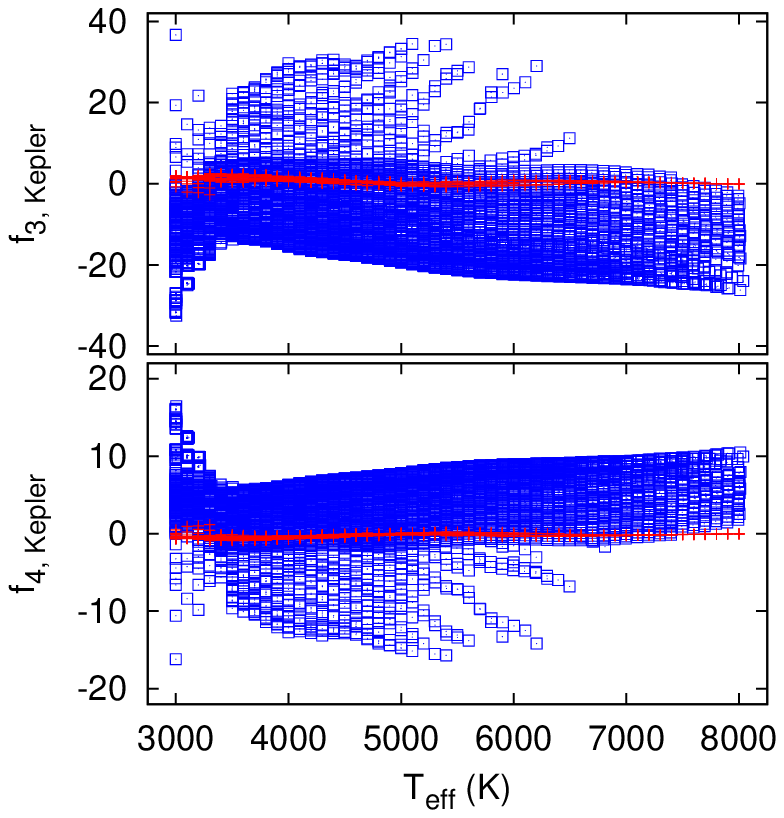}
\end{center}
\caption{Limb-darkening coefficients $f_1$, $f_2$, $f_3$ and $f_4$ used in the 
\citet{Claret2000} four-parameter law, Eq.~\ref{eq:4-p}, applied to the \textit{Kepler} 
 photometric band.  The symbols are the same as in Fig.~\ref{fig:ab_l}. }
\label{fig:ab_4p}
\end{figure*}

We present in Figs.~\ref{fig:ab_l}, \ref{fig:ab_q}, \ref{fig:ab_log} 
and \ref{fig:ab_4p} the coefficients derived by least-squares fitting for the limb-darkening 
laws given by Eqs.~1 -- 6 respectively for the 
{\it Kepler} photometric band              as a function of 
effective temperature  for both plane-parallel and 
spherically symmetric model stellar atmospheres.               It is 
clear that more realistic spherically symmetric model stellar atmospheres predict 
limb-darkening coefficients that vary much more as a function of effective temperature than those for plane-parallel models.  For the 
simplest case of the linear limb-darkening law, the $u$-coefficient determined 
from plane-parallel models in the {\it Kepler}-band vary from $u = 0.2$ 
to $0.5$, whereas spherical models with the same effective temperatures 
and gravities vary from $u = 0.6$ to $1.4$.  The coefficients predicted for 
almost all limb-darkening laws examined   here         show the same 
behavior as the limb-darkening coefficients predicted for a flux-conserving 
linear+square-root law \citep{Neilson2011, Neilson2012}.  This uniform dependence of the coefficients on $T_{\rm{eff}}$  is 
surprising and suggests all of these laws carry essentially the same information regarding the moments of the intensity and the atmospheric extension
about the stellar atmosphere in question.  The one exception is the 
\citet{Claret2000} four-parameter limb-darkening law, for which the 
coefficients appear to vary much more as a function of effective 
temperature. 

\begin{figure*}[t]
\begin{center}
\includegraphics[width=0.5\textwidth]{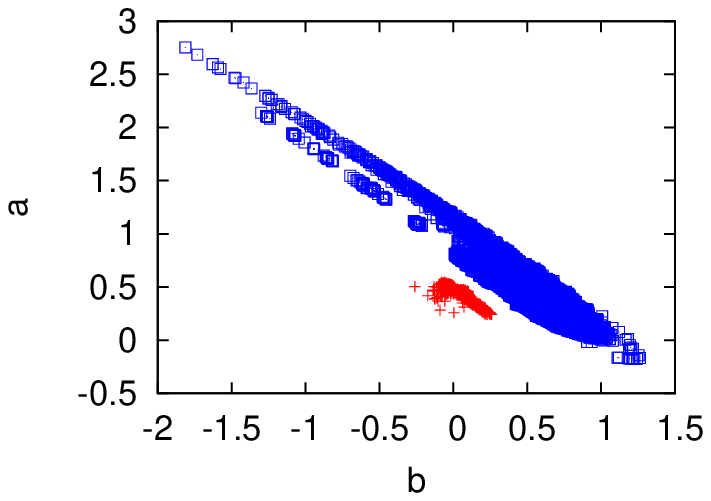}\includegraphics[width=0.5\textwidth]{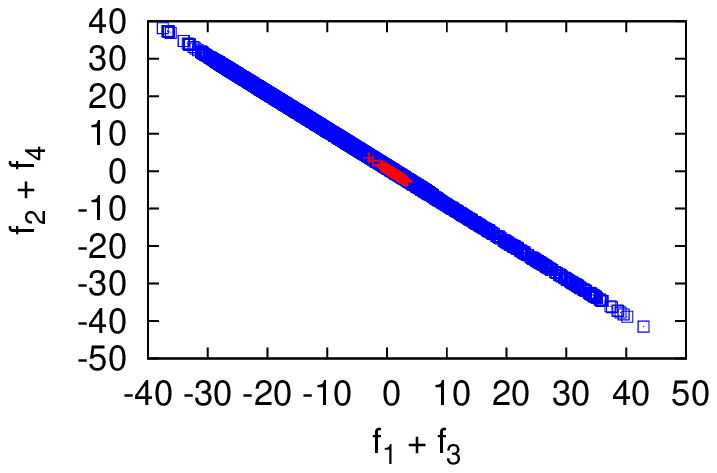}
\end{center}
\caption{Correlation between the {\it Kepler}-band limb-darkening 
coefficients for the quadratic law (left) and for the four-parameter law, $f_2 + f_4$ as a function of 
$f_1 + f_3$ (right). Red crosses represent coefficients computed from 
plane-parallel models and blue squares spherical models.}
\label{fig:4p_corr}
\end{figure*}

To explore the interdependence of the coefficients, we plot in Fig.~\ref{fig:4p_corr} the {\it Kepler}-band $b$-coefficient from the quadratic law as a function of the $a$-coefficient.  This plot is typical of all the 
two-parameter limb-darkening laws considered in this work as well as 
the limb-darkening law employed   by \citet{Neilson2011, Neilson2012}, 
including the apparent hook in the correlation between coefficients.  
 Fig.~\ref{fig:4p_corr} also plots the values of $f_2 + f_4$ as a 
 function of $f_1 + f_3$ for the four-parameter law, again for the \textit{Kepler} 
 photometric band.  The correlation for both plane-parallel and 
spherical models is readily apparent.  A best-fit linear relation to 
the coefficients for spherical models is
\begin{equation}\label{eq:4pcor}
f_{2,\rm{Kepler}} + f_{4,\rm{Kepler}} = -0.989 (f_{1,\rm{Kepler}} + f_{3,\rm{Kepler}}) + 1.051.
\end{equation}
The correlation is different for plane-parallel models for which the 
slope is $-0.978$ and the intercept is $0.493$.

These correlations are caused by               the limb-darkening 
coefficients being linear combinations of various angular moments of the intensity.
For instance, in plane-parallel model atmospheres the moments 
$J\equiv \int I(\mu)d\mu$ and $K\equiv \int I(\mu)\mu^2 d\mu$ are 
              related such that $J = 3K$ \citep{Mihalas1978}.  In 
spherical symmetry, this ratio varies, causing the moments of 
the intensity to differ in spherical symmetry from those predicted moments for 
plane-parallel model stellar atmospheres.  This difference in geometry is reflected 
in the difference between the zero-points of the relation 
Eq.~\ref{eq:4pcor} for spherical models and that for plane-parallel models.  One can 
potentially use this difference to test observations and test model 
geometry.

\subsection{Error Analysis}

\begin{figure*}[t]
\begin{center}
\includegraphics[width=0.5\textwidth]{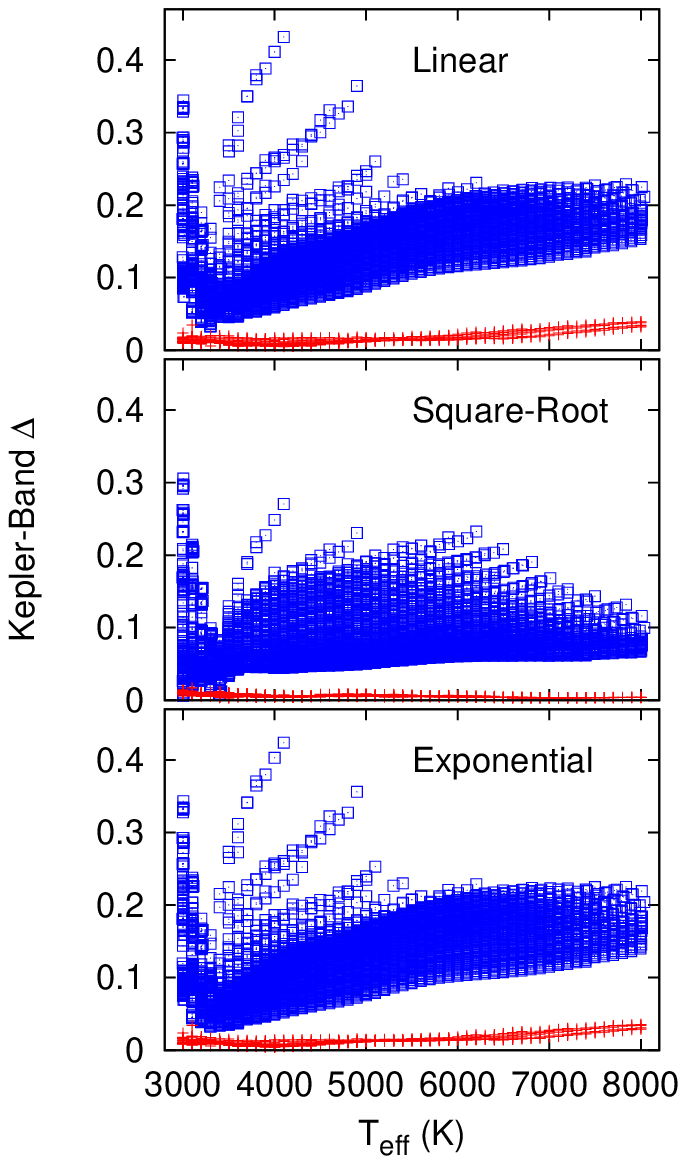}\includegraphics[width=0.5\textwidth]{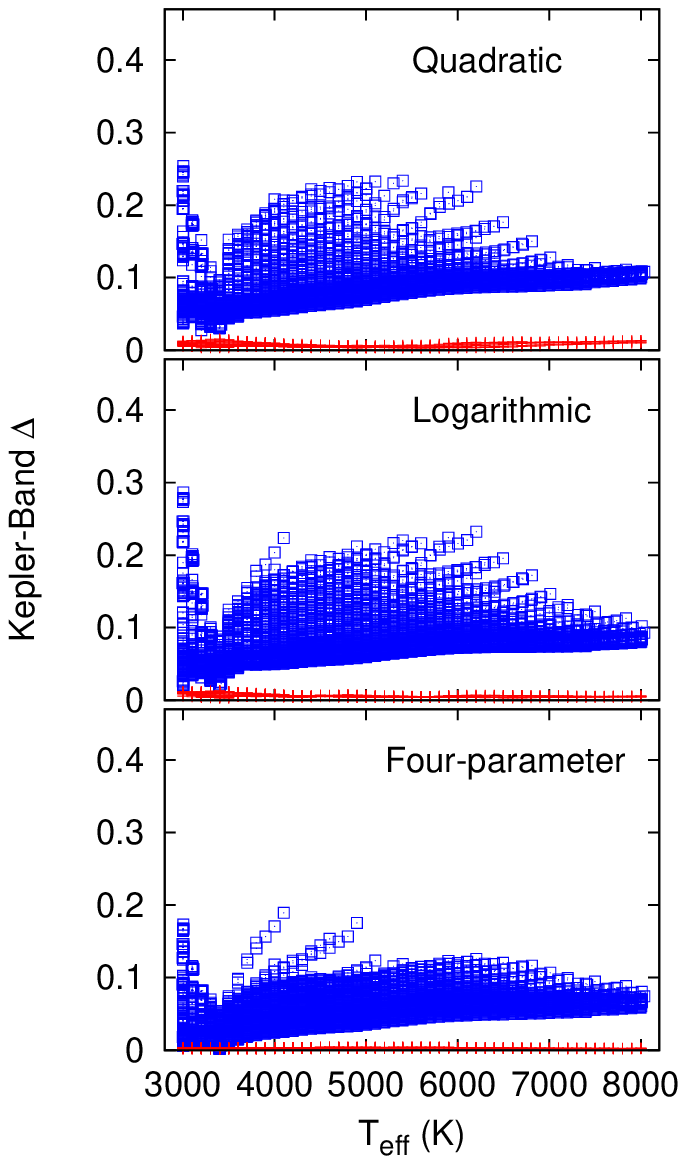}
\end{center}
\caption{The error of the best-fit limb-darkening relation, defined by 
Eq.~\ref{eq:delta}, for every model atmosphere (red crosses represent 
plane-parallel models, blue squares spherical models) for each of the 
six limb-darkening laws at {\it Kepler}-band wavelengths.}
\label{fig:err_l}
\end{figure*}

Various limb-darkening laws, such as those given in 
Eqs.~\ref{eq:linear}--\ref{eq:ln}, are fit to the surface intensities 
computed with model stellar atmospheres, and 
it is important to understand how well these laws represent the actual intensities.  For instance, 
\citet{Diaz1995} argued that a square-root law fit intensity profiles 
for hotter stars ($T_\mathrm{eff} > 9000~$K) better than a quadratic law, 
whereas no limb-darkening law is preferred for cooler stars.  We 
compute the relative error of the limb-darkening fit, $\Delta$, using 
the relation
\begin{equation}\label{eq:delta}
\Delta_\lambda \equiv \sqrt{ \frac{\sum [I_\mathrm{model}(\mu) - I_\mathrm{fit}(\mu)]^2}{\sum [I_\mathrm{fit}(\mu)]^2}},
\end{equation}
which quantifies the deviation of the best-fit limb-darkening law 
                  from the surface intensities of the model atmosphere.  We 
compute the relative error for each bandpass as a function of the 
fundamental stellar parameters for both plane-parallel and spherical 
geometries, and show in Fig.~\ref{fig:err_l} the relative errors as a function of effective temperature 
for fits in the {\it Kepler}-band. The relative error of the fits for 
spherical models is greater than the error for plane-parallel fits for 
all the limb-darkening laws.  The errors are similar only for           
              $T_\mathrm{eff} \sim 3500~$K, where                  the 
spherical model atmospheres predict intensity profiles that are closest 
to being linear, with           the error of the linear limb-darkening 
law approaching a minimum value.  This result appears to 
suggest that these limb-darkening laws are inappropriate for fitting 
light curves and interferometric observations, but this is not true. 

There are a number of issues with how the relative error is computed 
and what the error tells us, such as how the limb-darkening laws are 
defined, how they are fit to the surface intensities and the effect of 
sampling.  

\begin{itemize}
\item \textbf{Defining limb-darkening laws:}
The intensity profiles computed using the plane-parallel and spherical model 
atmospheres employed in this work are normalized with respect to the central 
intensity so that $I(\mu = 1) \equiv 1$.  Furthermore, all limb-darkening laws, except 
the exponential law, are defined so that the $I(\mu =1) \equiv 1$, regardless 
of the values of the best-fit coefficients.  As a result,     every fit 
to an intensity profile is anchored to the center of the stellar disk 
before representing the remainder of the intensity profile.  This definition alone results 
in a perfect fit to the center of the stellar disk and a deteriorating fit as 
$\mu \rightarrow 0$ as the                  intensity profile deviates 
from the assumed structure of a particular limb-darkening law.

\item \textbf{Fitting limb-darkening laws:}
Limb-darkening laws are typically fit to intensity profiles using a 
least-square         algorithm.  \citet{Neilson2011} showed that the  
                   best-fit coefficients for a given law are functions of weighted 
integrals of the intensity profile.  For example, the linear 
limb-darkening coefficient from Eq.~1 is a function of the mean 
intensity, $J$, and the stellar flux, 
$\mathcal{H} \equiv \int I(\mu)\mu d\mu$, and both of these quantities are more 
sensitive to the central intensity than to the much smaller intensity 
near the limb.  As with the definition of the limb-darkening laws, 
using a least square fitting algorithm fits the central part of the 
intensity structure better.
Similarly, one might fit limb-darkening coefficients by enforcing flux 
conservation, but because the flux is the $\mu$-weighted integral of 
the intensity,             any flux-conserving fit is constrained 
weakly by the intensity at the stellar limb relative to the intensity 
near the center of the stellar disk.

\item \textbf{Sampling Issues:}
Sampling is the most important of the three issues affecting the 
computed error of the fit of the intensity profile.  For instance,  \citet{Wade1985} and \citet{Heyrovsky2007} noted that fitting an 
intensity profile that is uniformly sampled in $\mu$ has a larger              
error than fitting the same profile that is uniformly sampled in 
$r = \sin\theta = \sin(\cos^{-1}\mu)$.  Uniform $r$-spacing emphasizes the 
intensity profile near the center of the disk while a uniform $\mu$-spacing emphasizes 
the limb.   Adopting any of the limb-darkening laws presented here, that
law will fit the central part of the stellar surface more precisely 
than the limb because of the normalization at the center of the disk. 
If, in addition, the surface intensity is sampled uniformly in $r$, 
that will give added weight to the central region.  These two factor 
combine to make the computed error of the fit smaller.  
Similarly, \citet{Howarth2011} found that limb-darkening coefficients 
derived  from planetary transits with large impact factors do not agree 
with model stellar atmosphere predictions.  This is because           
      the planet passes across only the limb of the star and not the 
center, therefore probing only part of the intensity profiles.   \cite{Claret2008, Claret2009} also found disagreement between theoretical limb-darkening coefficients and empirical coefficients measured from eclipsing binary light curves and comparisons to the planetary system HD 209458. 
Limb-darkening coefficients from stellar atmosphere models fit the 
whole profile yielding different results.
\end{itemize}

The combination of these three factors lead to calculated errors that 
are relative and not an absolute measure of the quality of the fit.  In 
this work, differences in the error between fits to plane-parallel and 
spherically symmetric model stellar atmosphere intensity profiles computed with the same 
properties are due solely to differences in the intensity profile near the 
limb where the spherical models provide more realistic predictions.  
Therefore, the error analysis suggests that the various limb-darkening 
laws lack the necessary complexity to precisely fit intensity profiles from spherical models.  The only 
exception is the \citet{Claret2000} four-parameter law, which fits the 
laws best, but appears to have unique properties.

\section{Gravity Darkening Coefficients}

Rapid rotation distorts the shape of a star, making it        
aspheric, with flattened poles and a bulged               
            equator. As shown first by \citet{vonZeipel1924}, the gravity and effective temperature    
vary in a coordinated way across the stellar surface 
such that at any point                                the effective 
temperature                                        is proportional to 
the effective gravity,               $T_\mathrm{eff} \sim g_\mathrm{eff}^{\beta_1/4}$,
where $\beta_1 = 1$ for radiative stars.  However, this value of 
$\beta_1$ is valid only for bolometric radiation, and \citet{Kopal1959} later 
derived monochromatic gravity-darkening corrections, $y(\lambda)$.  
\citet{Claret2000}, \citet{Claret2003} and \citet{Claret2011} have 
computed waveband-dependent gravity-darkening corrections as a 
        a function of the central intensity of the star, as well as the 
gravity, effective temperature and the variable, $\beta_1$ from plane-parallel models.
\citet{Bloemen2011} derived
\begin{equation} \label{gd:eq}
y(\lambda) = \left(\frac{\partial \ln I(\lambda)}
                        {\partial \ln g}\right)_{T_\mathrm{eff}} + 
             \left(\frac{\mathrm{d}\ln T_\mathrm{eff}}
                        {\mathrm{d}\ln g} \right ) 
             \left(\frac{\partial \ln I(\lambda)}
                        {\partial \ln T_\mathrm{eff} } \right)_g,
\end{equation}
and noted that 
$(\mathrm{d} \ln T_\mathrm{eff}/\mathrm{d} \ln g) = \beta_1/4$.   The variable $\beta_1$ is a function of effective temperature, but for the purpose of this analysis we assume $\beta_1 = 0.2$ for $T_{\rm{eff}} < 7500$~K and  $\beta_1 = 1$  for hotter stars. However, the value of $\beta_1$ based on  von Zeipel's theorem is not strictly valid for radiative or convective stellar envelopes \citep{Claret2000b, Espinosa2011, Claret2012c}. 

\begin{figure*}[t]
\begin{center}
\includegraphics[width=0.5\textwidth]{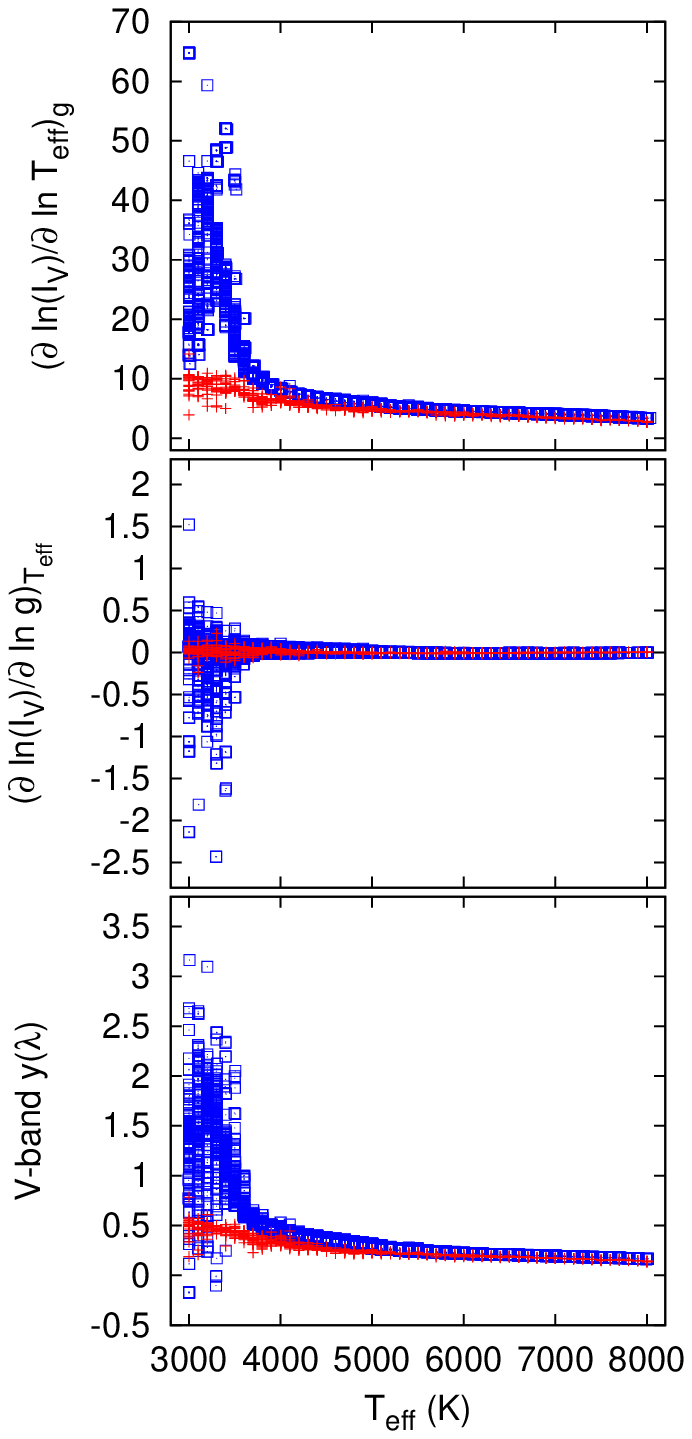}\includegraphics[width=0.5\textwidth]{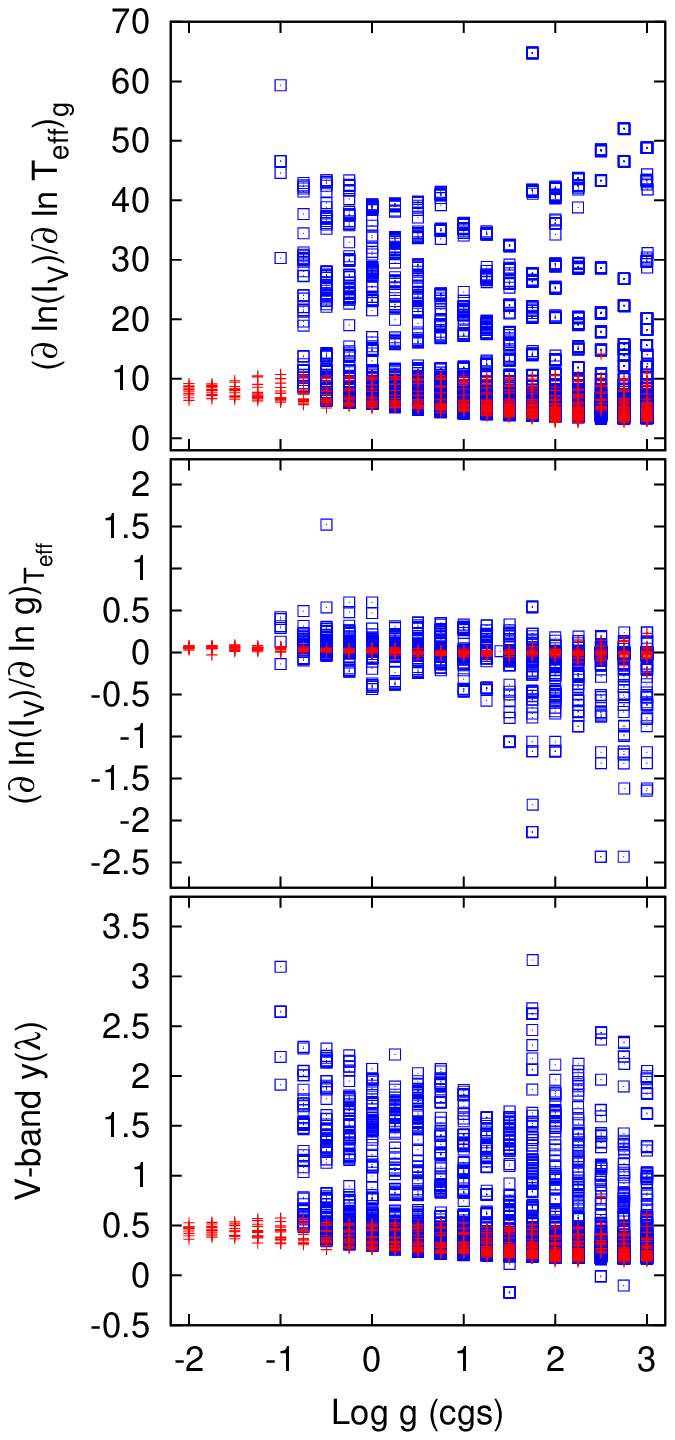}
\end{center}
\caption{$V$-band central intensity derivatives and gravity-darkening 
coefficients as function of effective temperature (left) and gravity 
(right) computed from plane-parallel (red crosses) and 
spherically symmetric (blue squares) model stellar atmospheres.}
\label{fig:gd}
\end{figure*}

In Fig.~\ref{fig:gd}, we plot     the $V$-band     values of each intensity derivative for 
each model stellar atmosphere in Eq.~\ref{gd:eq}, as well as $y(\lambda)$ computed for the assumed values of $\beta_1$.  We 
find that plane-parallel and spherically symmetric model stellar 
atmospheres predict similar gravity-darkening coefficients for 
$T_\mathrm{eff} > 4000~$K, but there are significant 
differences for cooler stars.  We interpret these differences for the cooler stars as consequences of both surface 
convection and the shift from the negative hydrogen ion to titanium 
oxide as the dominant opacity source.                             Both 
plane-parallel and spherical model intensities show greater variation 
at these cool effective temperatures, but the intensity profiles   of 
spherically symmetric model atmospheres vary more than that of 
plane-parallel model atmospheres.
 
While the most significant differences between spherical and planar 
model predictions of gravity-darkening coefficients are at lower 
temperatures, the gravity-darkening coefficients computed from 
spherically symmetric models are greater than those of plane-parallel 
models for every effective temperature. For example, a  
spherically symmetric model with $T_\mathrm{eff} = 8000~$K has a $V$-band 
gravity-darkening coefficient of $y_V \simeq 0.165$ while the   
plane-parallel model with the same effective temperature has 
$y_V \simeq 0.14$.  The difference is small but systematic.

\section{Angular Diameter Corrections}

Interferometric observations measure the angular diameter of a 
star along with its limb-darkening profile, but, unfortunately, the measured 
angular diameter and limb-darkening profiles are not independent 
quantities.  This is especially true when the measured visibilities do not probe the second lobe.  \citet{Davis2000} measured 
stellar angular diameters from interferometric observations by assuming 
that the stellar intensity profile is uniform, \textit{i.e.} the intensity at 
any point on a stellar disk is equal to the central intensity.  In that 
case, the uniform-disk angular diameter can be directly fit to the 
observed visibilities and then converted to a limb-darkened angular 
diameter using model stellar atmospheres. \citet{Davis2000} computed 
corrections using plane-parallel \textsc{Atlas} models \citep{Kurucz1993} and found
$k \equiv \theta_\mathrm{UD}/\theta_\mathrm{LD} = 0.91$ to $0.98$ in the wavelength 
range $\lambda = 400$-$800~$nm.  These limb-darkening corrections have 
been applied to observations of Cepheids \citep{Gallenne2012} and 
Sirius \citep{Davis2011} for example.

\begin{figure*}[t]
\begin{center}
\includegraphics[width=0.5\textwidth]{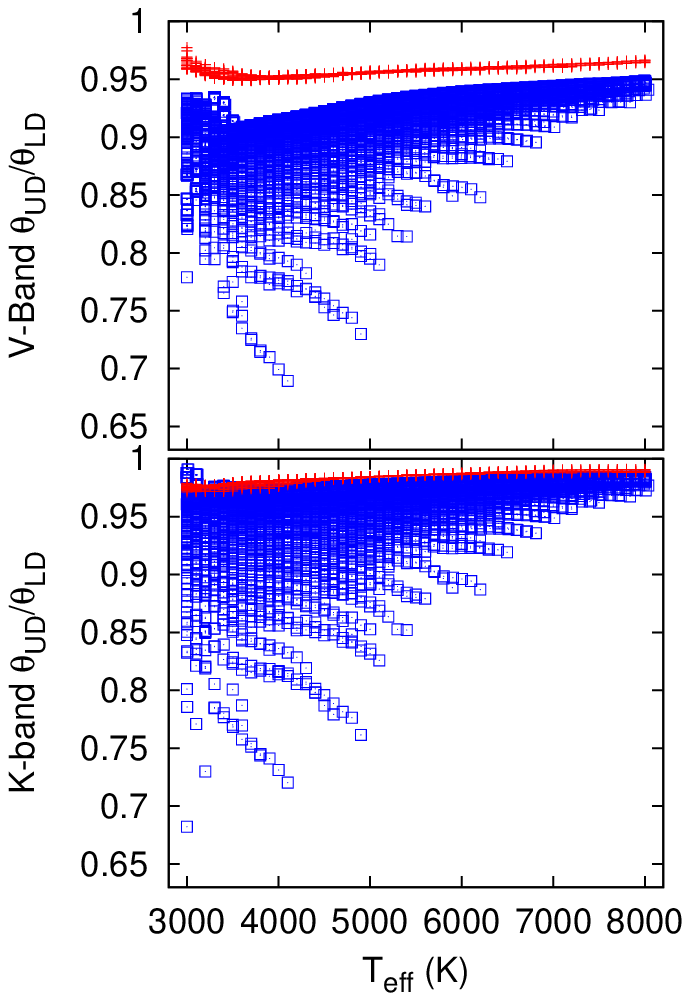}\includegraphics[width=0.5\textwidth]{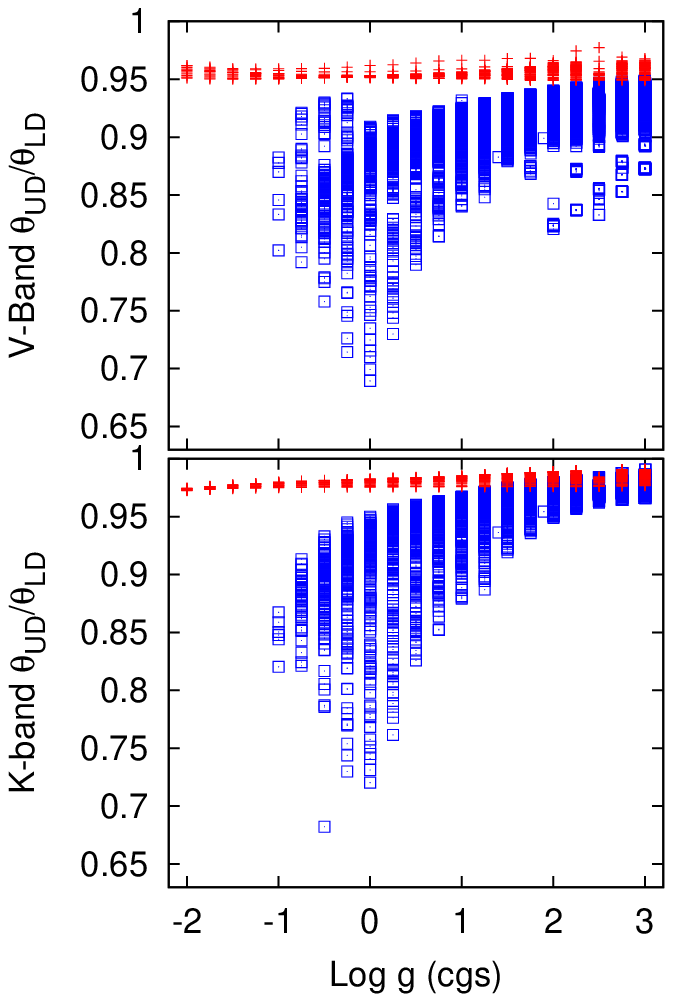}
\end{center}
\caption{Interferometric angular diameter correction computed in 
$V$-band (top) and $K$-band as functions of effective temperature (left)
and gravity (right). Corrections computed from plane-parallel model 
atmospheres are denoted with red x's and spherically symmetric models 
blue squares.}
\label{fig:ang}
\end{figure*}

We compute angular diameter corrections using the recipe described by 
\citet{Marengo2004}, where we assume a limb-darkened  angular 
diameter of $\theta_\mathrm{LD} = 1~$mas to compute interferometric visibilities from a 
model atmosphere intensity profile.              That synthetic visibility is then 
fit by  a uniform-disk angular diameter.  The best-fit uniform-disk 
angular diameter is then equivalent to the theoretical angular diameter 
correction.  We compute angular diameter corrections for the 
Johnson-Cousins $BVRIHK$ wavebands and show the 
corrections for the $V$- and $K$-bands in Fig.~\ref{fig:ang} as a function of 
effective temperature for plane-parallel and spherically symmetric 
models.  Corrections from spherical models clearly differ from 
corrections from plane-parallel model atmospheres. Intensity profiles 
from plane-parallel model stellar atmospheres predict corrections in 
the narrow range from $k = 0.97$ - $0.99$ in $V$-band and approaches unity for 
longer wavelengths.  We show in Fig.~\ref{fig:ang} the $V$- and $K$-band 
angular diameter corrections as function of effective temperature and 
gravity for plane-parallel and spherically symmetric model stellar 
atmospheres.

Fits to spherically symmetric model atmospheres suggest significantly 
different angular diameter corrections as functions of both effective 
temperature and gravity.  The $V$-band corrections from spherical 
models, denoted $k_s$, range from $k_s = 0.67$ to $0.95$, with no 
overlap with the plane-parallel model predictions.  The $K$-band corrections 
show similar behaviors except that spherical and planar corrections 
overlap somewhat.  These results suggest that using plane-parallel 
model atmosphere corrections systematically underestimates the 
stellar angular diameter.  For instance, \citet{Mozurkewich2003} 
presented uniform-disk angular diameters for a sample of 85 stars, 
along with limb-darkened angular diameters corrected using 
limb-darkening coefficients from \citet{Claret1995} and \citet{Diaz1995}. 
Their angular diameter corrections vary from $k = 0.89$ to $\approx 1$, 
consistent with the values  found here for plane-parallel model atmospheres. 

Of particular interest are the results of \citet{Mozurkewich2003} for 
$\alpha$ Persei (F5\,Ib), for which they measured 
$T_\mathrm{eff} = 6750~$K, and for $\epsilon$ Geminorum (G8\,Ib), which 
was measured to have $T_\mathrm{eff} = 4485~$K.  
\citet{Mozurkewich2003} measured the uniform-disk angular diameters at 
$550~$nm to be $2.986\pm0.042$ for $\alpha$ Per and $4.467\pm 0.115$~mas for $\epsilon$ Gem. Using these they computed limb-darkened angular diameters of $3.188\pm0.035$ and 
$4.703\pm0.047$~mas, respectively. 
Our spherically-symmetric models with $\log g = 1.5$ and 
$M = 10~M_\odot$ yield $V$-band angular-diameter corrections 
of $0.929$ for $\alpha$ Per and $ 0.916$ for $\epsilon$ Gem.  Applying these to the uniform disk measurements 
gives    larger limb-darkened angular diameters: 
$\theta_\mathrm{LD} = 3.214$~mas for $\alpha$ Per and 
$\theta_\mathrm{LD} = 4.877$~mas for $\epsilon$ Gem.
                             The spherical correction  for $\alpha$ Per 
yields a value for $\theta_\mathrm{LD}$ that is marginally consistent with the 
angular diameter found using plane-parallel correction, whereas the 
limb-darkened angular diameter of $\epsilon$ Gem measured by 
\citet{Mozurkewich2003} is almost $4\%$ smaller than what would be 
predicted by applying spherical model corrections.  This difference may 
appear to be small but this underestimate is systematic.

\begin{figure}[t]
\begin{center}
\includegraphics[width=0.5\textwidth]{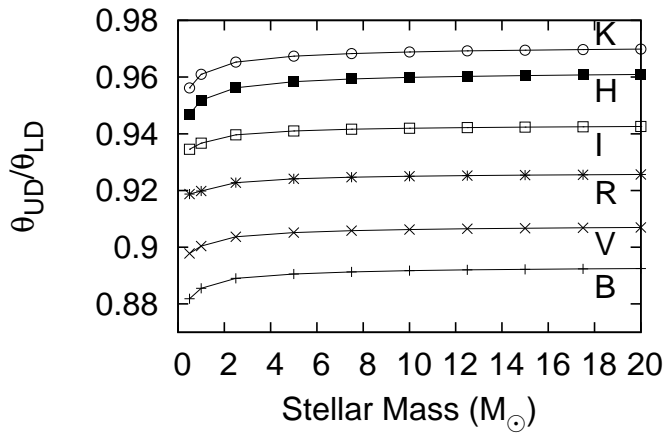}
\end{center}
\caption{Interferometric angular diameter corrections as a function of 
waveband and stellar mass for spherically symmetric model stellar 
atmospheres with $\log g = 2$ and $T_\mathrm{eff} = 3500$~K.}
\label{fig:ang_mass}
\end{figure}

As a test, we check how the angular diameter corrections vary as 
function of stellar mass. Because models with low effective temperature 
but relatively high gravity appear to predict the smallest corrections, 
we hold $T_\mathrm{eff} = 3500$~K and $\log g = 2$.
             The angular diameter 
corrections are shown in Fig.~\ref{fig:ang_mass} as a function of 
stellar mass for the six Johnson-Cousins wavebands considered in this 
work.  The figure suggests that the corrections are insensitive to the 
mass of the stellar model except for low-mass ($M \le 1~M_\odot$) 
models. This is reassuring and suggests that when applying these 
corrections, one can ignore the stellar mass.  The difference between limb-darkening profiles and angular diameter corrections is small and consistent with previous results by \cite{Lester2013}.

\section{Summary}

In this work, we present model atmosphere intensity profiles for the 
$BVRIHK$,     {\it CoRot} and {\it Kepler} passbands from    both 
plane-parallel and spherically symmetric geometries based on models 
computed by \citet{Neilson2011, Neilson2012}.  We fit a number of 
limb-darkening laws to these intensity profiles, as well as compute 
gravity-darkening coefficients and angular diameter corrections for 
interferometry.  We test how these fits vary as a function of model 
atmosphere geometry and compile tables of limb-darkening coefficients, 
gravity-darkening coefficients and angular diameter corrections that 
can be applied to observations.

We consider six limb-darkening laws in this work: linear, quadratic, 
square-root, four-parameter, exponential and logarithmic.  These laws 
fit the intensity profiles from plane-parallel model atmospheres well, but 
not the intensity profiles of the spherical models based on computed relative errors. The one exception  
is the \citet{Claret2000} four-parameter law, for which 
the difference between the spherical model intensity profiles and the 
predictions of the fitting law is small enough to still be 
applicable to observations, although the law still fits the spherical 
profiles more poorly than the plane-parallel intensities.

While those predicted errors are useful for comparing fits to planar 
and spherical model intensity profiles, they are not ideal for studies  
of actual  limb darkening.  Best-fit limb-darkening coefficients depend 
on the definition of the laws, all of which anchor the fit to 
$I(\mu = 1) = 1$, making              the fit    sensitive to the 
sampling of the intensity profile as well as to the method for fitting the 
data.  Because intensity profiles for spherical models are more complex,
     the fitting error is greater than the error for simpler plane-parallel 
model intensity profiles. However, spherically symmetric model 
atmospheres are a more realistic representation of actual stellar 
atmospheres, meaning they are better suited for limb darkening studies.

Fits to the four-parameter limb-darkening law also show correlations 
between the limb-darkening coefficients; we find that the linear 
combination of the four coefficients are approximately constant, with  
that constant being a function of the atmosphere's geometry.  This 
result suggests that the linear combination of the observed 
coefficients for the four-parameter law provides a simple test of 
whether the observations are probing the edge of the stellar disk, 
\textit{i.e.} sphericity.

We also predict wavelength-dependent gravity-darkening coefficients 
based of the \citet{Claret2011} prescription.  Unlike the limb-darkening
coefficients, the gravity-darkening coefficients are less dependent on 
model atmosphere geometry.  This is because the gravity-darkening 
coefficients depend on the change of the central intensity with respect 
to effective temperature and gravity, hence the difference between 
atmospheres for the same geometry. Gravity-darkening is also a function of the central intensity, which is insensitive to model geometry.  The spherically symmetric 
gravity-darkening coefficients are similar to plane-parallel 
coefficients for $T_\mathrm{eff} > 5000~$K and begin to diverge for 
cooler stellar atmosphere models.  Only at the coolest effective 
temperatures, $3000~$K$ \le T_\mathrm{eff} \le 4000~$K, is the geometry 
of the model atmosphere important, with the spherically symmetric coefficients 
being approximately an order-of-magnitude greater than those predicted 
from plane-parallel model atmospheres.

Unlike the gravity darkening coefficients, the interferometric 
angular-diameter corrections do depend on geometry.  
For plane-parallel model atmospheres the angular-diameter corrections  
vary from about 
$0.95$ -- $1$, whereas the corrections for spherically symmetric model 
atmospheres vary from $0.67$ -- $1$.                      Previous 
analyses had assumed that corrections from plane-parallel models
are applicable to all stars, but this is not true. At low gravity, 
$\log g < 3$, spherically symmetric corrections deviate significantly 
from plane-parallel model predictions.  The difference between 
spherical and plane-parallel models is a function of both gravity and 
effective temperature and also appears to vary as a function of 
stellar mass.

The angular-diameter corrections, limb-darkening and 
gravity-darkening coefficients are publicly available as online tables. 
Each table has the format $T_\mathrm{eff}~$(K) , $\log g$, $M~(M_\odot)$, 
$R~(R_\odot)$ and $L~(L_\odot)$ and then the appropriate variables for 
each waveband, such as linear limb-darkening coefficients. Tables of 
gravity-darkening coefficients also contain values of the intensity 
derivatives with respect to gravity and effective temperature. For 
plane-parallel models, values of mass, radius and luminosity are 
presented as zero in the tables. We list the properties of these tables 
in Tab.~\ref{t1}, that are archived in electronic form at the CDS.  Model atmosphere intensity profiles are also archived at the CDS

\begin{table}[t]
\caption{Summary of limb-darkening coefficient, gravity-darkening 
coefficient and interferometric angular diameter correction tables 
found online.}\label{t1}
\begin{center}
\begin{tabular}{lll}
\hline
\hline
Name & Geometry & Type \\
\hline
Table2  & Spherical & Linear Limb Darkening Eq.~\ref{eq:linear} \\
Table3  & Spherical & Quadratic Limb Darkening Eq.~\ref{eq:quad} \\
Table4  & Spherical & Square Root  Limb Darkening Eq.~\ref{eq:root} \\
Table5  & Spherical & Four-parameter Limb Darkening Eq.~\ref{eq:4-p} \\
Table6  & Spherical & Exponential Limb Darkening Eq.~\ref{eq:exp} \\
Table7  & Spherical & Logarithmic Limb Darkening Eq.~\ref{eq:ln} \\
Table8  & Planar    & Linear Limb Darkening Eq.~\ref{eq:linear} \\
Table9  & Planar    & Quadratic Limb Darkening Eq.~\ref{eq:quad} \\
Table10  & Planar    & Square Root  Limb Darkening Eq.~\ref{eq:root} \\
Table11 & Planar    & Four-parameter Limb Darkening Eq.~\ref{eq:4-p} \\
Table12 & Planar    & Exponential Limb Darkening Eq.~\ref{eq:exp} \\
Table13 & Planar    & Logarithmic Limb Darkening Eq.~\ref{eq:ln} \\
Table14 & Spherical & Gravity Darkening \\
Table15 & Planar    & Gravity Darkening \\
Table16 & Spherical & Angular Diameter Corrections \\
Table17 & Planar & Angular Diameter Corrections \\
\hline
\end{tabular}
\end{center}
\note{Tables listed here can be retrieved electronically from the CDS.}
\end{table}

Techniques such as optical interferometry, microlensing observations, 
planetary transit and eclipsing binary observations are continuously improving 
the measurements of stellar limb darkening needed to test model stellar 
atmospheres and the physics assumed in their calculation.  At lower gravities, these 
observations      require the more physically realistic 
spherically symmetric models to constrain stellar properties.  The 
predicted limb-darkening coefficients, gravity-darkening coefficients 
and angular diameter corrections from spherically symmetric \textsc{SAtlas} 
models are new tools that for aiding analyses of these observations.
\acknowledgements

The authors acknowledge support from a research grant from the Natural 
Sciences and Engineering Research Council of Canada, the Alexander von 
Humboldt Foundation and National Science Foundation (AST-0807664).

\bibliographystyle{aa} 

\bibliography{ld3}

\end{document}